\documentclass[12pt]{iopart}
\usepackage{iopams}
\usepackage{graphicx}

\usepackage{color}
\usepackage{cite}

\begin{document}
\title{Interacting particles in an activity landscape}
\author{Adam Wysocki$^1$, Anil K. Dasanna$^1$, Heiko Rieger$^{1,2}$}
\address{$^1$ Department of Theoretical Physics and Center for Biophysics, Saarland University, 66123 Saarbr\"ucken, Germany}
\address{$^2$ INM – Leibniz Institute for New Materials, Campus D2 2, 66123 Saarbr\"ucken, Germany}

\ead{\mailto{a.wysocki@lusi.uni-sb.de}, \mailto{anilkumar.dasanna@uni-saarland.de}, \mailto{heiko.rieger@uni-saarland.de}}

\begin{abstract}
  We study interacting active Brownian particles (ABPs) with a space-dependent swim velocity via simulation and theory. We find that, although an equation of state exists, a mechanical equilibrium does not apply to ABPs in activity landscapes. The pressure difference originates in the flux of polar order and the gradient of swim velocity across the interface between regions of different activity. In contrast to motility-induced phase separation of ABPs with a homogeneous swim velocity, a critical point does not exist for an active-passive patch system, which continuously splits into a dense and a dilute phase with increasing activity. However, if the global density is so high that not all particles can be packed onto the inactive patch, then MIPS-like behaviour is restored and the pressure is balanced again.
\end{abstract}
\submitto{\NJP}	


        
\section{Introduction}

Active particles, like motile microorganisms, live in a complex environment and are affected by various external fields like gravity, fluid flows or walls \cite{bechinger2016RMP}. Mostly it is assumed that their intrinsic properties, like propulsion speed or tumbling rate, are constant in space \cite{shaebani2020NRP}. This is generally not true because organisms often respond to a spatially varying stimulus such as light intensity or chemical concentration. For example, the bacteria {\it Escherichia coli} are chemotactic: they sense concentration gradients and adjust their tumbling rate to swim up or down a chemical gradient. Others do not respond to gradients but react in a non-directional way to the local stimulus intensity, a behaviour called kinesis. For instance, some cyanobacteria are photokinetic and their swim speed depends on the local light intensity \cite{wild2017FEMS}. In this context, it has been predicted theoretically that the local density of particles $\rho(\mathbf{r})$ performing a run-and-tumble motion is inversely proportional to their local propulsion speed $v(\mathbf{r})$ \cite{schnitzler1993PRE}. This fact was used to arrange millions of light-powered bacteria into a complex pattern, such as Leonardo da Vinci's {\it Mona Lisa}, via light fields \cite{frangipane2018eLife,arlt2018NC}. Optical fields can also be used to create activity landscapes for synthetic microswimmers, such as thermophoretic Janus particles \cite{soker2021PRL}. An activity landscape can have a technological application, such as traps \cite{jahanshahi2020CP}, but it can also be used to study fundamental questions of active matter \cite{auschra2021PRE}. One such question is whether there is a coexistence criterion for active matter. The introduction of an active component of pressure \cite{takatori2014PRL,winkler2015SM}, often called ``swim pressure'', allowed to formulate a mechanical equilibrium condition of equal pressures in phase separating purely repulsive active Brownian particles (ABPs) with a homogeneous propulsion speed \cite{solon2015NP}. The separation into a dense-inactive and a dilute-active phase is triggered by a slowdown during collisions and is named motility-induced phase separation (MIPS) \cite{cates2015ARCMP}. The behaviour of ABPs in activity landscapes, like the aggregation of particles in slow regions, resembles in some sense MIPS, however, the differences and similarities are not well studied. Here we show that an activity pattern can still lead to pressure imbalance between regions of different activity even though an equation of state exists for ABPs \cite{solon2015NP}. We show that, in contrast to MIPS, a critical point does not exist for an active-passive patch system, which continuously splits into a dense and a dilute phase with increasing activity \cite{hasnain2017SM,fischer2020PRE}. However, if the global density is so high that not all particles can be packed onto the inactive patch, then MIPS-like behaviour is restored and the pressure is balanced again. The particles in the low activity region exhibit all states of matter of a two dimensional system: a liquid, a hexatic phase and a crystal. The transition to a state with a high orientational order for a sufficiently dense and active system can be predicted theoretically.

\section{Model}

We consider $N$ active Brownian particles (ABPs) in 2D with a space-dependent propulsion velocity $v(\mathbf{r})$. The particles interact via a repulsive pair potential $V(r)=\frac{k}{2}(\sigma-r)^2$ if $r\leq \sigma$, i.e., the inter-particle distance $r$ is smaller then the particle diameter $\sigma=2a$, and $V(r)=0$ otherwise \cite{fily2014SM}. The repulsion strength $k$ is chosen such that the particle overlap is around $0.01\sigma$. The positions $\mathbf{r}_i=(x_i,y_i)$ and the orientations $\mathbf{e}_i=(\cos{\theta}_i,\sin{\theta_i})$ evolve according to the overdamped Langevin equations:  
\begin{eqnarray}
\dot{\mathbf{r}}_i&=&v(\mathbf{r}_i)\mathbf{e}_i+\mu_t\sum_{j\neq i}\mathbf{f}_{ij}+\sqrt{2D_t}\,\boldsymbol{\eta}_i\label{eom_pos} \label{eqn1}\\
\dot{\theta}_i&=&\sqrt{2D_r}\,\xi_i\,,\label{eom_angle}
\end{eqnarray}
where $1/\mu_t=\gamma_t$ is the translational friction coefficient, $\mathbf{f}_{ij}=\mathbf{f}(\mathbf{r}_i-\mathbf{r}_j)=-\nabla_{\mathbf{r}_i}V(|\mathbf{r}_i-\mathbf{r}_j|)$ is the force on $i$-th particle due to $j$-th particle, $D_t$ and $D_r$ are the translational and rotational diffusion constant and $\boldsymbol{\eta}_i$, $\xi_i$ are zero-mean unit-variance Gaussian white noises. For spherical Brownian particles it is $D_r = 3 D_t/\sigma^2$. We use a simulation box of size $L_x/\sigma = 160$ and $L_y/\sigma = 40$ with periodic boundary conditions in both directions and consider mainly a step-like activity landscape 
\begin{equation}
    v(x)=v_a\,\Theta\left(\alpha L_x/2-|x|\right)\,,
\end{equation}
where $\Theta$ denotes the unit step function and $\alpha$ controls the extent of the active region ($\alpha=0.5$ means that a half of the box is active). Our length and time unit is the particle radius $a$ and the rotational diffusion time $\tau_r=1/D_r$, respectively. Thus, the P\'eclet number, $Pe=v_a/(a D_r)$, is a dimensionless measure of activity and $\phi_0=\pi a^2 N/(L_x L_y)$ is the global packing fraction.

\section{Theory}
First we ask whether a mechanical equilibrium exists or in other words: are pressures equal in neighboring regions of different mobility? 

Starting from (\ref{eom_pos}) and (\ref{eom_angle}) one obtains the full Smoluchowski equation \cite{solon2015NP,solon2015PRL,solon2018NJP,paliwal2018NJP}
\begin{equation}
  \partial_t\psi=D_t\nabla^2\psi+D_r\partial_{\theta}^2\psi-\nabla\cdot\left[v(\mathbf{r})\mathbf{e}\psi+\int\mathrm{d}\mathbf{r}'\,\mu_t\mathbf{f}(\mathbf{r}'-\mathbf{r})\left\langle\hat{\rho}(\mathbf{r}')\hat{\psi}(\mathbf{r},\theta)\right\rangle\right]
    \label{smoluchowski}
\end{equation}
for the time evolution of the noise-averaged probability density $\psi(\mathbf{r},\theta)=\langle\hat{\psi}\rangle=\langle\sum_{i=1}^N\delta(\mathbf{r}-\mathbf{r}_i)\,\delta(\theta-\theta_i)\rangle$, where $\langle\dots\rangle$ denotes noise averages and $\hat{\rho}=\int\mathrm{d}\theta\,\hat{\psi}$ the fluctuating particle density. In the following we assume that the system is in the steady state, $\partial_t\psi=0$, and is only inhomogeneous along the $x$-direction. Integrating (\ref{smoluchowski}) over $\theta$ gives
\begin{equation}
    0=\partial_x J_{\rho}(x)
    \label{density1}
\end{equation}
with the particle flux 
\begin{equation}
J_{\rho}(x)=v(x)m(x)+I_1(x)-D_t\partial_x\rho(x)\,,
    \label{density2}
\end{equation}
where $m=\int\mathrm{d}\theta\,\cos(\theta)\psi$ is the polarisation and $I_1=\int\mathrm{d}\mathbf{r}'\,\mu_t f_x(\mathbf{r}'-\mathbf{r})\left\langle\hat{\rho}(\mathbf{r}')\hat{\rho}(\mathbf{r})\right\rangle$ is a contribution due to the pair-potential. Similarly, multiplying (\ref{smoluchowski}) by $\cos(\theta)$ and integrating over $\theta$ gives
\begin{equation}
      D_rm(x)=-\partial_x J_m(x)
    \label{polar1}
\end{equation}
with the flux of polar order
\begin{equation}
     J_m(x)=v(x)\left\{\frac{\rho(x)}{2}+Q(x)\right\}+I_2(x)-D_t\partial_x m(x)\,,
    \label{polar2}
\end{equation}
where $Q=\frac{1}{2}\int\mathrm{d}\theta\,\cos(2\theta)\psi$ encodes nematic order along $x$-direction and $I_2=\int\mathrm{d}\mathbf{r}'\,\mu_t f_x(\mathbf{r}'-\mathbf{r})\left\langle\hat{\rho}(\mathbf{r}')\hat{m}(\mathbf{r})\right\rangle$ contains the effect of interactions. All setups considered here are flux-free steady states, i.e., $J_{\rho}(x)=0$. Thus inserting (\ref{polar1}) into the vanishing particle flux $J_{\rho}$ in (\ref{density2}) yields
\begin{equation}
    0=-\frac{v}{D_r}\partial_x J_m+I_1-D_t\partial_x\rho\,.
    \label{flux}
\end{equation}
Let us next consider two neighboring domains $A$ and $B$ with constant activity $v_A$ and $v_B$, respectively. Then multiplying (\ref{flux}) with $\gamma_t=k_\mathrm{B}T/D_t$, integrating from the bulk of region $A$ to the bulk of region $B$ and using integration by parts gives
\begin{equation}
     \int_A^B\mathrm{d}x\,\left(k_\mathrm{B}T\partial_x\rho-\gamma_t I_1\right)=-\left[\frac{\gamma_t}{D_r}v J_m\right]_A^B+\frac{\gamma_t}{D_r}\int_A^B\mathrm{d}x\,J_m\partial_xv\,.
\end{equation}
We define the ideal passive pressure as $p_{id}=\rho k_\mathrm{B}T$, the interaction or ``direct'' pressure as $\partial_x p_D=-\gamma_t I_1$ \cite{irving1950JCP,solon2015PRL} and the ``swim'' pressure as $p_S=\gamma_t v^2\rho/2D_r+\gamma_t v I_2/D_r$ \cite{takatori2014PRL,yang2014SM,solon2015PRL}. For a homogeneous bulk phase in region $A$ and $B$ we can set $Q=\partial_x m=0$ and obtain finally
\begin{equation}
    \left[p_{id}+p_D+p_S\right]_A^B=\frac{\gamma_t}{D_r}\int_A^B\mathrm{d}x\,J_m\partial_xv\,,
    \label{pressure_diff_general}
\end{equation}
which is the main result of our study.\\ 
Especially, for an activity step, where an ABP swims with $v_A$ for $x<x_{if}$ and abruptly propels with $v_B$ when crossing the interface at $x=x_{if}$, we get $\partial_x v=(v_B-v_A)\delta(x-x_{if})$ and
\begin{equation}
    \left[p_{id}+p_D+p_S\right]_A^B=\frac{\gamma_t}{D_r}J_m(x_{if})(v_B-v_A)\,.
    \label{pressure_diff_step}
\end{equation}
What does this mean? It means that the bulk pressures in both activity regions are not equal and that the pressure difference is governed by the activity profile $v(x_{if})$ and the flux of polar order $J_m(x_{if})$ at the interface between this regions. Thus an inhomogeneous activity pattern can still lead to unequal pressures in both phases even though an equation of state exists for ABPs \cite{solon2015NP}. A similar expression as (\ref{pressure_diff_general}) have been obtained for underdamped quorum-sensing active particles undergoing motility-induced phase separation \cite{fily2017JPA}, which lack an equation of state for the pressure even for spatially uniform activity \cite{solon2015NP}.

Next, we examine the case of non-interacting ABP's, which can be solved exactly in all details \cite{auschra2021PRE,row2020PRE}. In particular, let us consider a setup where particles are inactive ($v_p=0$) for $x<x_{if}=0$ and are active ($v_a>0$) for $x>0$. We non-dimensionalize (\ref{pressure_diff_step}) using $\pi a^2/k_\mathrm{B}T$. The left hand side of (\ref{pressure_diff_step}) reads as 
\begin{equation}
    \frac{\pi a^2}{k_\mathrm{B}T}\left[p_{id}+p_S\right]_p^a=\phi_a\left(1+\frac{v_a^2}{2D_t D_r}\right)-\phi_p=\left(\frac{\lambda_p}{\lambda_a}-1\right)\phi_p\,
    \label{lhs_ideal}
\end{equation}
using the bulk density ratio $\lambda_a/\lambda_p=\rho_a/\rho_p$ \cite{magiera2015PRE,hasnain2017SM,row2020PRE,fischer2020PRE,auschra2021PRE}, 
where $\lambda_{\alpha}=\left(D_r/D_t+v_{\alpha}^2/2D_t^2\right)^{-1/2}$ denotes the polarisation decay length and $\phi=\pi a^2\rho$ the packing fraction. In the following we use the polarisation profile obtained in \cite{auschra2021PRE,row2020PRE}, which in the passive region ($x<0$) takes the form $m(x)=m_{\mathrm{max}}\exp{(x/\lambda_p)}$ with $m_\mathrm{max}=-v_a\rho_p\lambda_a\lambda_p/2D_t(\lambda_a+\lambda_p)$ for our particular setup. In order to calculate the right hand side of Eq.~(\ref{pressure_diff_step}) we use 
\begin{equation}
    J_m(0)=\int_{-\infty}^0\mathrm{d}x\,\partial_x J_m(x)=-D_r\int_{-\infty}^0\mathrm{d}x\,m(x)=-D_r m_\mathrm{max}\lambda_p\,,
\end{equation}
where $J_m(-\infty)=0$ because of $v(-\infty)=\partial_xm(-\infty)=0$, see (\ref{polar2}) for the individual terms of $J_m$. In total we get for the right hand side of (\ref{pressure_diff_step})
\begin{equation}
   \frac{\pi a^2}{k_\mathrm{B}T}\frac{\gamma_t}{D_r} J_m(0)v_a=\left(\frac{\lambda_p}{\lambda_a}-1\right)\phi_p>0\,,
   \label{rhs_ideal}
\end{equation}
 which is equivalent to (\ref{lhs_ideal}) and thus proves the validity of (\ref{pressure_diff_step}) for the ideal system. Due to $\lambda_p>\lambda_a$, the pressure in the active region is larger than the pressure in the passive region. 

With (\ref{pressure_diff_step}) we want to estimate the densities $\phi_a$ and $\phi_p$ of interacting particles in an active-passive patch system. An approximate equation of state for interacting ABP's in 2D \cite{takatori2015PRE} reads as
\begin{equation}
    p_a=p_{id}+p_S+p_D=\rho_a k_\mathrm{B}T\left[1+\frac{3}{8}Pe^2\left(1-\frac{\phi_a}{\phi_\mathrm{max}}\right)+\frac{3}{2\pi}\frac{Pe\,\phi_a}{1-\frac{\phi_a}{\phi_\mathrm{max}}}\right]\,,
    \label{EOS_active}
\end{equation}
where $\phi_\mathrm{max}=\pi/\sqrt{12}$ is the packing fraction of a close-packing of monodisperse disks. The ``swim'' pressure vanishes and the interaction pressure diverges at $\phi_\mathrm{max}$, respectively. For passive hard disks we use \cite{santos1995JCP}
\begin{equation}
    p_p=\rho_p k_\mathrm{B}T\left(1-2\phi_p+\frac{2\phi_\mathrm{max}-1}{\phi_\mathrm{max}^2}\phi_p^2\right)^{-1}\,,
    \label{EOS_passive}
\end{equation}
 which, however, does not capture the existence of a ``hexatic'' phase at $\phi\in[0.7,0.72]$, see \cite{bernard2011PRL}. We obtain one equation by inserting (\ref{EOS_active}) and (\ref{EOS_passive}) into (\ref{pressure_diff_step}) using the interface contribution of non-interacting particles (\ref{rhs_ideal}), however, with a density dependent activity $v(\phi)=v_a(1-\phi/\phi_\mathrm{max})$. And we get another equation by assuming a rectangular density profile such that $\phi_0=\alpha\phi_a+(1-\alpha)\phi_p$. The numerical solution of these two equations give the densities $\phi_a$ and $\phi_p$ for each $Pe$ and $\phi_0$.

\section{Results}

\begin{figure}[th!]
  \centering
	\includegraphics[scale=0.82]{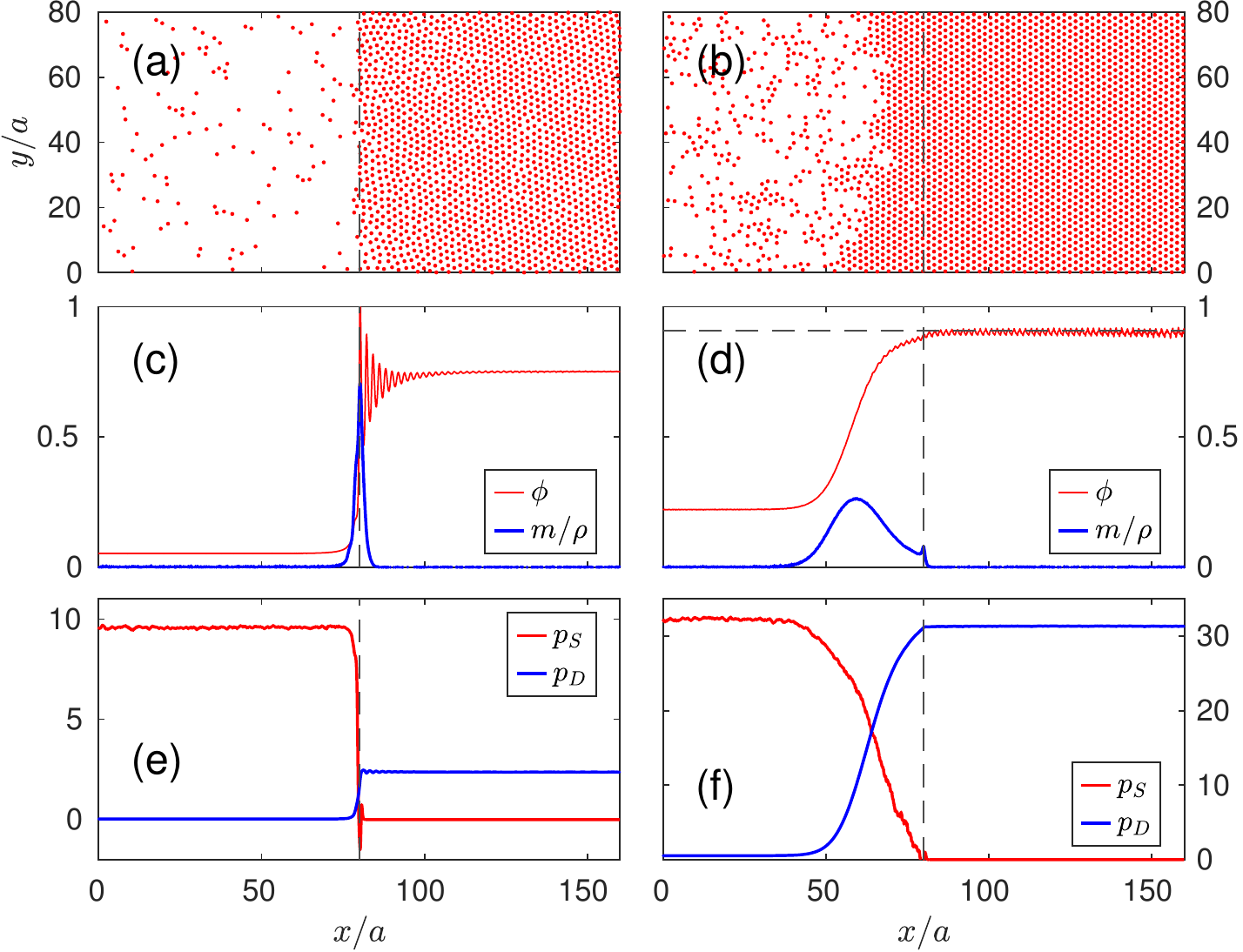}
	\caption{(a,b) Snapshots of active Brownian particles in an active-passive patch system along with the corresponding (c,d) packing fraction $\phi(x)$ (red) and polarization profiles $m(x)/\rho(x)$ (blue) and (e,f) the local ``swim'' $p_S(x)$ (red) and ``direct'' pressure profiles $p_D(x)$ (blue) for two different global packing fractions $\phi_0=0.4$ (left column) and $\phi_0=0.65$ (right column) but for the same P\'eclet number $Pe=40$ are shown. Only half of the simulation box is shown $[0,L_x/2]$ and the dashed vertical line at $x=L_x/4$ indicates the interface between the active (left) and the passive region (right), i.e., the fraction of the active region is $\alpha=0.5$.}
	\label{fig1}
\end{figure}

Figure~\ref{fig1}(a,b) shows snapshots of a system with a step-like activity pattern with $\alpha=0.5$, i.e., half of the box, namely $[-L_x/4,L_x/4]$, is active and the rest is passive. In Figure~\ref{fig1}(a,c,e) the global packing fraction is $\phi_0=0.4<(1-\alpha)\,\phi_\mathrm{max}\approx 0.4534$ and the maximum possible packing fraction in the passive patch is below close-packing. Figure~\ref{fig1}(a) demonstrates the general behaviour of a system with a position dependent activity $v(\mathrm{r})$, namely, that particles tend to accumulate in the less mobile region or in other words, $\rho\propto 1/v$ \cite{schnitzler1993PRE,tailleur2008PRL}. A strong positive polarization $m$ (particles point towards the passive region) appears solely at the active-passive interface and is zero otherwise, see figure~\ref{fig1}(c). Similarly to the behavior of ABP's near walls \cite{elgeti2013EPL}, only the particles that point towards the interface can also approach it and once they cross the interface they keep their orientation for $\tau_r$. In Figure~\ref{fig1}(e) we show the normal component of the local ``swim'' $p_S(x)$ \cite{das2019SR} and interaction pressure tensor $p_D(x)$ \cite{todd1995pre}. Surprisingly, the pressure is not equal in both regions, even though an equation of state for ABPs exists \cite{solon2015NP}, and is larger in the dilute active region, where the ``swim'' pressure $p_S$ is the dominant contribution. There is a pressure gradient, but we do not observe particle flux. How to resolve this contradiction? One can take two points of view. If one argues that the ``true'' pressure consists only of the ideal $p_{id}$ and the ``direct'' pressure $p_D$ (neglecting the ``swim'' pressure $p_S$) then the pressure gradient is balanced by the swim force density $\gamma_t vm$ created by the polarization of the active particles \cite{omar2020PRE} or in other words
\begin{equation}
\partial_x\left(p_{id}+p_D\right)=\gamma_t v(x)m(x)\,,
\end{equation}
which follows from $J_{\rho}(x)=0$. For our case that means that the dense passive phase is held together by the rim of polarized particles in the active region. If one also considers $p_S$ then the pressure gradient is balanced by the term on the right hand side of (\ref{pressure_diff_general}) consisting of the flux of polar order $J_m$ and the gradient in the activity $\partial_x v$. For a dense system, $\phi_0>(1-\alpha)\,\phi_\mathrm{max}$, the situation is different and is reminiscent of so-called motility-induced phase separation \cite{fily2014SM,takatori2015PRE,solon2015PRL}, as can be seen in figure~\ref{fig1}(b,d,f). The dense phase extends into the active region and the polarized region is no longer located at the active-passive boundary but at the interface between the dense and the dilute phase. Moreover, this interface fluctuates heavily and the pressure is constant through the system.

\begin{figure}[th!]
  \centering
	\includegraphics[scale=.525]{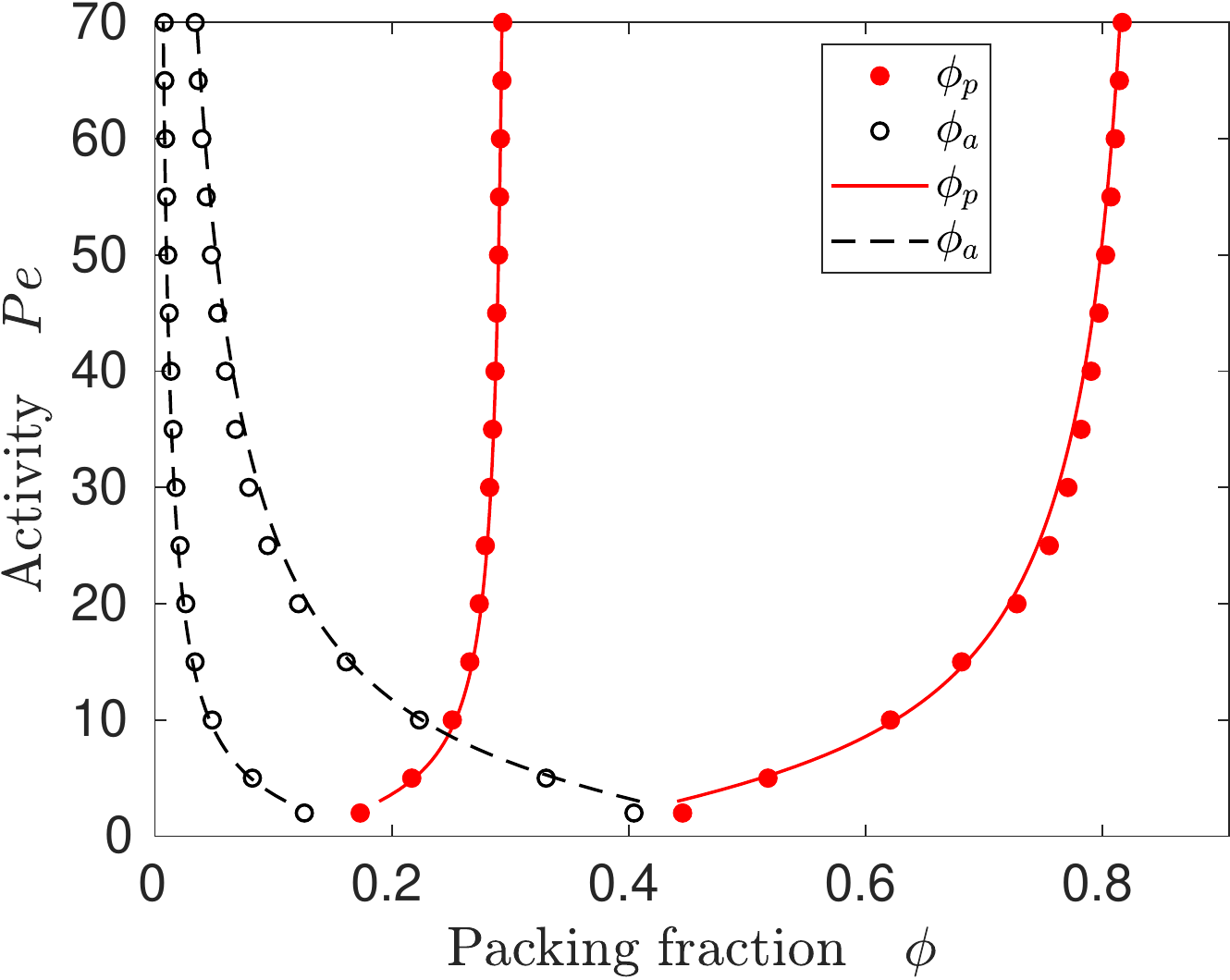}
	\caption{Packing fractions in the active and the passive region, $\phi_a$ and $\phi_p$, as a function of the activity $Pe$ for two different global packing fractions $\phi_0=0.15$ and $\phi_0=0.425$, below overcrowding, $\phi_0<(1-\alpha)\,\phi_\mathrm{max}$ with $\alpha=0.5$, obtained from simulations (dots) and theory (lines).}
	\label{fig2}
\end{figure}

ABPs with a homogeneous propulsion speed separate only above a critical point $Pe_{\rm cr}\approx 26.7$, $\phi_{\rm cr}\approx 0.597$ into a dense-immobile and a dilute-mobile phase \cite{siebert2018PRE}. The densities of the coexisting phases are independent of the global density $\phi_0$ at fixed activity $Pe$. The situation is completely different for systems with a mobility pattern. In figure~\ref{fig2} the packing fractions $\phi_a$ and $\phi_p$ as a function of $Pe$ for two different $\phi_0$ below overcrowding, $\phi_0<(1-\alpha)\,\phi_\mathrm{max}\approx 0.4534$, are shown. No signs of a critical point are visible, i.e., the system splits continuously, starting at $Pe=0$, into a dense and a dilute phase with increasing $Pe$. In contrast to the ideal case, the densities $\phi_a$ and $\phi_p$ depend on the global density $\phi_0$ at fixed $Pe$ and in the limit $Pe\to\infty$ all particles occupy the passive patch, $\phi_p=\phi_0/(1-\alpha)$ and $\phi_a=0$, as long as $\phi_0<(1-\alpha)\,\phi_\mathrm{max}$. Moreover, figure~\ref{fig2} demonstrates that the theory, based on (\ref{pressure_diff_step}), agree very well with the simulation results for densities below overcrowding.

\begin{figure}[th!]
  \centering
	\includegraphics[scale=0.55]{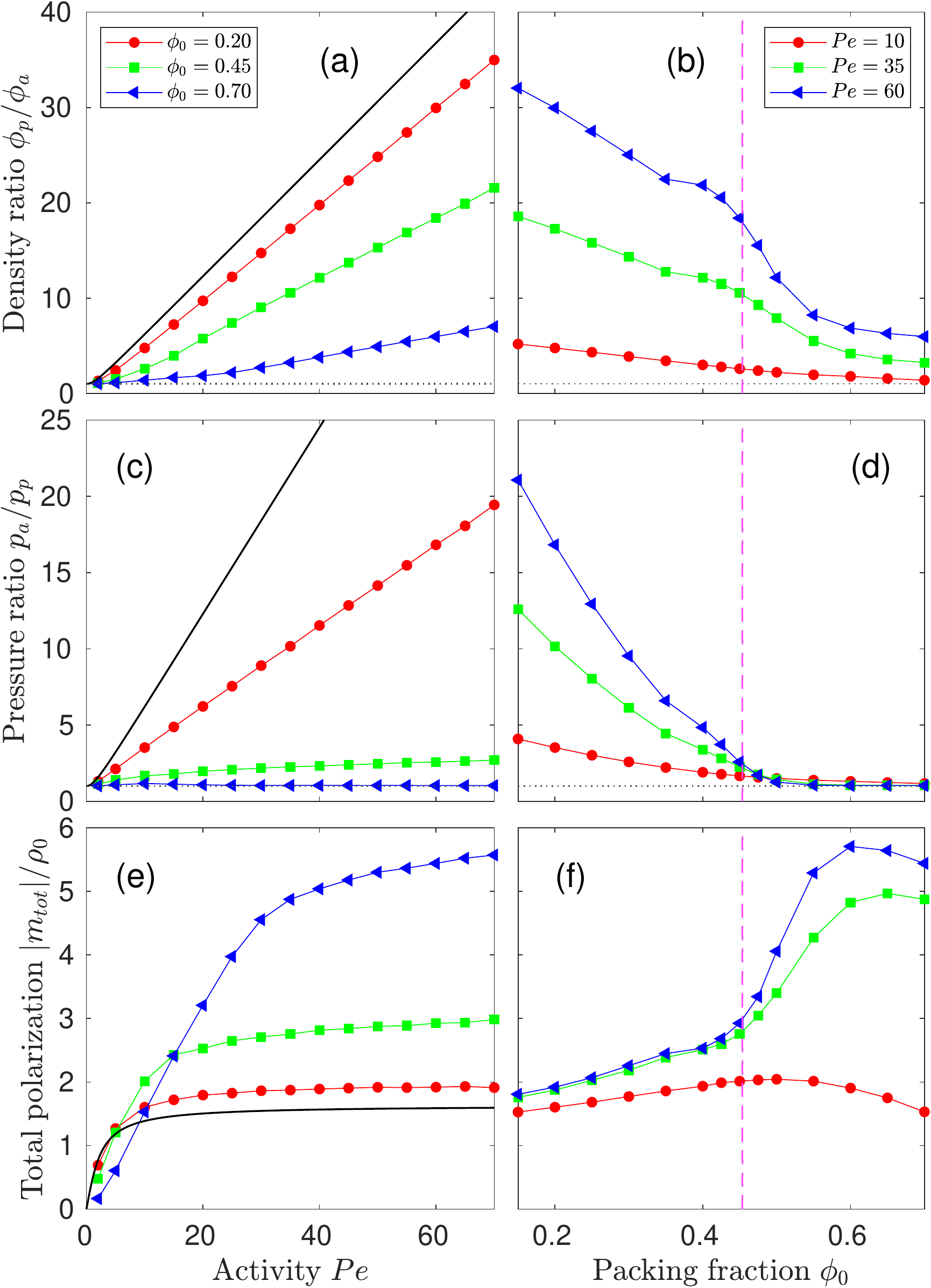}
	\caption{(a,b) Density ratio $\phi_p/\phi_a$, (c,d) pressure ratio $p_a/p_p$ and (e,f) the total interface polarization $m_{\rm tot}=\int_a^p\mathrm{d}x\,m(x)$ for varying activities $Pe$ and global packing fractions $\phi_0$ are shown. In (a,c,e) the solid black lines indicate the exact results for the non-interacting case. The dotted horizontal line in (a-d) is drawn at $1$, i.e, when both passive and active densities or pressures become equal. The vertical dashed line in (b,d,f) is drawn at $\phi_0=(1-\alpha)\,\phi_{\mathrm{max}}$ with $\alpha=0.5$ and indicates the global packing fractions $\phi_0$ above which the passive patch could be completely filled up in the limit $Pe\to\infty$.}
	\label{fig3}
\end{figure}

\begin{figure}[th!]
  \centering
	\includegraphics[scale=0.45]{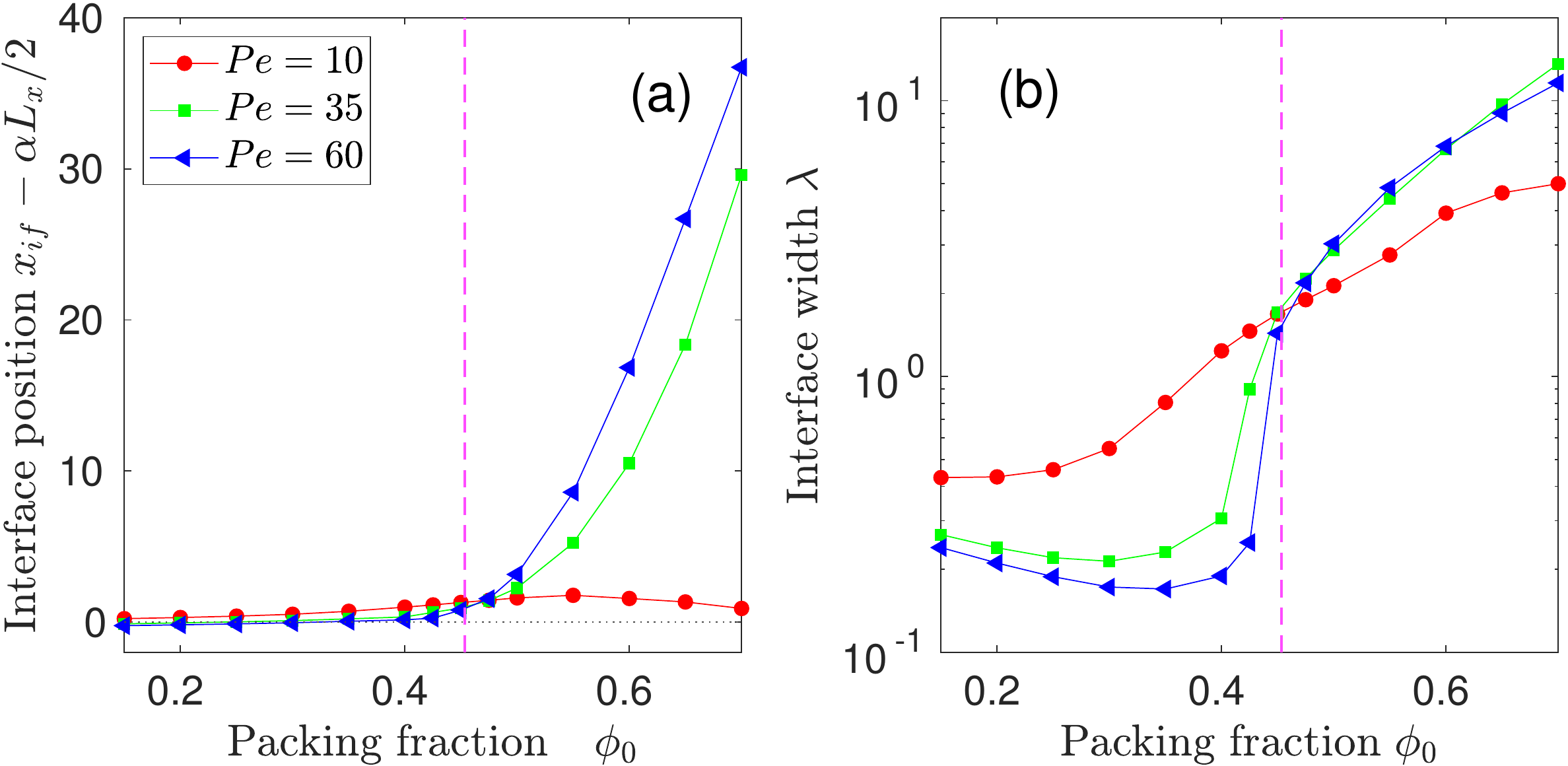}
	\caption{(a) Interface position $x_{if}$ and the interface width $\lambda$ as function of the overall packing fraction $\phi_0$ for various activities $Pe$. The vertical dashed line is drawn at $\phi_0=(1-\alpha)\phi_{\mathrm{max}}$ with $\alpha=0.5$, which marks the border of overcrowding in the limit $Pe\to\infty$.}
	\label{fig4}
\end{figure}

Figure~\ref{fig3} indicates the interdependencies between the densities of the coexisting phases, the corresponding pressures and the polarisation at the interface between both phases. Independent of the global density, the density ratio behaves as for ideal particles, $\phi_p/\phi_a\propto Pe$, see figure~\ref{fig3}(a), however, $\phi_p/\phi_a$ decreases with $\phi_0$ at fixed $Pe$ due to excluded volume effects as is evident from figure~\ref{fig3}(b). The pressure ratio behaves similarly to the non-interacting case, $p_p/p_a\propto 1/Pe$, for $\phi_0<(1-\alpha)\,\phi_\mathrm{max}\approx 0.4534$, see figure~\ref{fig3}(c). However, the pressure in both phases becomes equal, $p_a=p_p$, for the overcrowded case, $\phi_0>(1-\alpha)\,\phi_\mathrm{max}$, see figure~\ref{fig3}(d). The total interface polarisation $m_\mathrm{tot}=\int_a^p\mathrm{d}x\,m(x)$, which is proportional to the flux of polar order $J_m$ in the bulk of the active region \cite{hermann2020PRR}, saturates with increasing $Pe$, see figure~\ref{fig3}(e). Besides the trivial dependency $m_\mathrm{tot}\propto\phi_0$ there is a pronounced change in growth of $m_\mathrm{tot}$ as a function of $\phi_0$ at $(1-\alpha)\,\phi_\mathrm{max}$ for large $Pe$, see figure~\ref{fig3}(f). Once the passive patch is fully occupied the dense phase have to extend into the active region, which means that the position of the interface between the dense and the dilute phase $x_{if}$ does not coincide anymore with the boundary of the active and passive patch $\alpha L_x/2$, see figure~\ref{fig4}(a). Consequently, the interface start to fluctuate strongly and its width $\lambda$ increases significantly above $(1-\alpha)\,\phi_\mathrm{max}$, see figure~\ref{fig4}(b), resembling the behaviour of an interface during motility-induced phase separation \cite{patch2018SM}. We obtained $x_{if}$ and $\lambda$ by fitting the density profile with a hyperbolic tangent.

\begin{figure}[th!]
  \centering
	\includegraphics[scale=0.27]{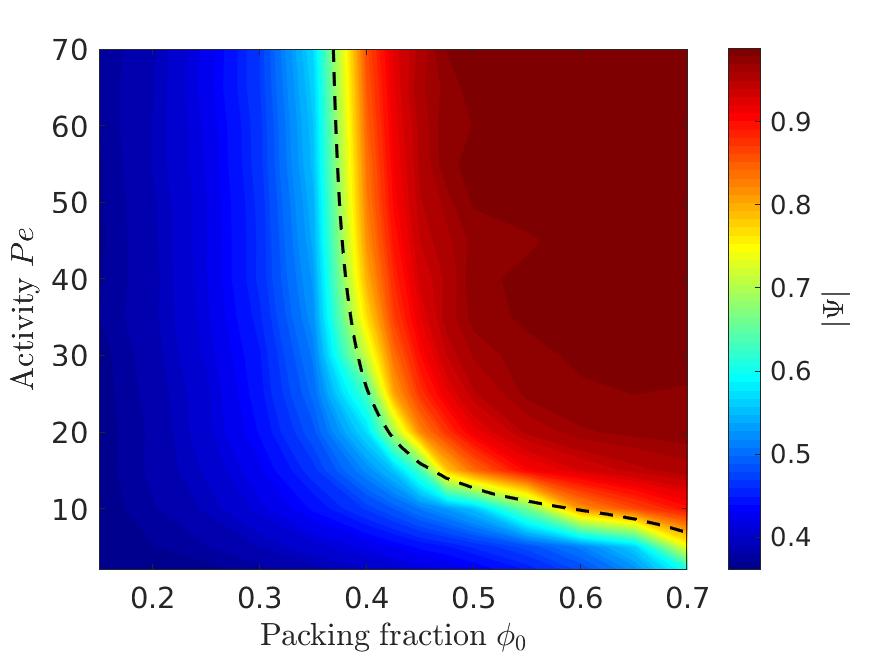}
	\caption{Average magnitude of the local bond-orientational order parameter $|\Psi|$ within the passive region as function of the global packing fraction $\phi_0$ and activity $Pe$. For a perfect triangular lattice $|\Psi|=1$. The isoline $\phi_p(\phi_0,Pe)=0.716$ (black dashed line) indicates the transition to a pure hexatic phase \cite{bernard2011PRL} and is a numerical result of the theory based on (\ref{pressure_diff_step}).}
	\label{fig5}
\end{figure}

As clearly visible in figure~\ref{fig1}(a), the less active patch has a much higher particle density than the other. However, monodisperse disks can only be packed up to $\phi_{\mathrm{max}}=\pi/\sqrt{12}$ arranged in a hexagonal lattice \cite{fejes1940MZ} and thus a crystalline order in the less mobile region is expected for a sufficiently dense and active system. Hexagonal packing has a long-range orientational (sixfold) order, which can be measured using the local hexatic order parameter $\Psi(\mathbf{r}_j)=\sum_{k=1}^{N_j}\exp{(\mathrm{i}6\theta_{jk})}/N_j$, where $\theta_{jk}$ is the angle formed by the bond that connects the $j$th disk and its $k$th (out of $N_j$) nearest neighbor (found with a Voronoi tessellation algorithm) and the $x$ axis \cite{bernard2011PRL}. For a perfect triangular lattice, all the angles $6\theta_{jk}$ are the same and $|\Psi(\mathbf{r}_j)|=1$. Because of the polycrystalline character of the dense phase we use the average magnitude of the hexatic order parameter $|\Psi|=\sum_{i=1}^N|\Psi(\mathbf{r}_i)|/N$ in order to get a global information on the order \cite{lotito2020ACIS}, which however does not vanish in an isotropic fluid. In figure~\ref{fig5} the order parameter $|\Psi|$ within the passive patch as function of $\phi_0$ and $Pe$ is shown. $|\Psi|$ displays a sharp increase beyond the isoline $\phi_p(\phi_0,Pe)=0.716$ (dashed line), which indicates the transition to a pure hexatic phase \cite{bernard2011PRL} and is a numerical result of the theory based on (\ref{pressure_diff_step}). The minimum global packing fraction necessary for a crystalline patch is $\phi_0=(1-\alpha)0.716$. 

\section{Summary}

 To summarize our main findings: we have theoretically justified the pressure imbalance in an activity landscape. The pressure difference originates in the flux of polar order and the gradient of swim velocity across the interface between regions of different activity. We found that although the density is lower in the more active area the corresponding pressure is higher. For not too dense systems, $\phi_0<(1-\alpha)\phi_{\mathrm{max}}$, the densities of the coexisting phases can be predicted from the momentum balance equation (\ref{pressure_diff_general}). Moreover, we have studied the effect of interactions on ABPs in activity landscapes. Excluded volume effect become significant above a global packing fraction $\phi_0=(1-\alpha)\phi_{\mathrm{max}}$, i.e., when, in the limit $Pe\to\infty$, the passive patch is completely filled with particles. In that case the system is reminiscent of a system with a homogeneous swim velocity undergoing MIPS and the pressure becomes equal in both phases. The dense phase in the less-active region exhibits a crystalline order for a sufficiently dense and active system, and the transition line can also be predicted from the momentum balance equation (\ref{pressure_diff_general}).

\ack
This work was financially supported by the German Research Foundation (DFG) within the Collaborative Research Center SFB 1027.

\section*{References}
\bibliographystyle{unsrt}
\bibliography{references}
\end{document}